\documentstyle[preprint,tighten,pra,aps,amssymb]{revtex}
\begin{document} \draft


\title{\LARGE \bf From Qubits to Continuous--Variable Quantum Computation}

\author{Barry C.\ Sanders}
\address{Department of Physics, Macquarie University, Sydney,
        New South Wales 2109, Australia}

\author{Stephen D.\ Bartlett}
\address{Department of Physics, Macquarie University, Sydney,
        New South Wales 2109, Australia}

\author{Hubert de Guise}
\address{Department of Physics, Lakehead University, Thunder Bay,
        Ontario, P7B 5E1, Canada}

\maketitle

\begin{abstract}
  By encoding a qudit in a harmonic oscillator and investigating the
  $d\to\infty$ limit, we give an entirely new realization of
  continuous--variable quantum computation.  The generalized Pauli
  group is generated by number and phase operators for harmonic
  oscillators.
\end{abstract}

\vspace{8mm}

The use of continuous--variable~(CV) quantum computing allows
information to be encoded and processed much more compactly and
efficiently than with discrete--variable~(qubit) computing.  With CV
realizations, one can perform quantum information processes using
fewer coupled quantum systems: a considerable advantage for the
experimental realization of quantum computing.  The rapidly developing
field of CV quantum information theory has applications to quantum
error correction~\cite{Got00}, quantum cryptography~\cite{Ral00} and
quantum teleportation~\cite{Bra98}, including an experimental
realization of CV quantum teleportation~\cite{Fur98}.

At present, the proposed realization of CV quantum computation employs
position eigenstates as a computational basis~\cite{Got00}; these
states are approximated experimentally using highly squeezed
states~\cite{Bra98}.  Here, we introduce new CV realizations, where
the generalized Pauli group is generated by the number operator
$\hat{N}$ and a phase operator $\hat{\theta}$, and the computational
basis is given either by harmonic oscillator number eigenstates or
phase eigenstates.  These realizations are obtained formally by taking
the $d \to \infty$ limit of the qudit, the $d$--dimensional
generalization of the qubit, and are important for four key reasons:
\begin{enumerate}
\item these CV realizations are entirely distinct from the position
  eigenstate computational basis realization, both in terms of the
  computational basis and in terms of the SUM gate;
\item the SUM gate employs a standard Kerr optical nonlinearity to
  couple two modes;
\item these realizations give natural extensions of the qubit--based
  (discrete--variable) Pauli group, with a well--defined limiting
  procedure; and
\item these realizations give a new implementation of the well--studied
  phase operator~\cite{Peg97}.
\end{enumerate}

In this letter, we review the generalized Pauli group for qudits, and
we construct the generators of the generalized Pauli group in $d$
dimensions using operators that will be shown to be expressible in
terms of the SU(2) angular momentum and phase operators.  This
construction allows us to conveniently view the $d$--dimensional space
of qudits as the Hilbert space of a $d$--dimensional irrep of SU(2).
We also express the qudits in terms of harmonic oscillator states and
investigate the $d\to\infty$ limit, and show that it is \emph{not} the
common generalization of the Pauli group for CV quantum information
(i.e., the Heisenberg--Weyl group) with position eigenstates as the
computational basis.  Instead, we obtain a new CV realization, where
the generalized Pauli group is generated by the number operator
$\hat{N}$ and a phase operator $\hat{\theta}$.  We also construct a
second realization of qudits in terms of phase states and show that
this realization is ``dual'' to the first realization given here.
Finally, we discuss a realization of CV quantum computation in coupled
harmonic oscillators.  We establish a SUM gate, which serves as the CV
analogue of the CNOT gate; this SUM gate employs a $\chi^{(3)}$
optical nonlinearity and is distinct in operation from the SUM gate
suggested~\cite{Got00} for the position--eigenstate computational
basis.

We begin by reviewing the Pauli group of a qubit, and its
generalization to the qudit.  A qubit is realized as a state in a
two--dimensional Hilbert space ${\cal H}_2$.  It is customary to
choose two normalized orthogonal states, $|0\rangle$ and $|1\rangle$,
to serve as a {\it computational basis} for ${\cal H}_2$.  The
unitary operators $\{ X_2 \equiv \sigma_x,\, Z_2 \equiv \sigma_z \}$,
where the $\sigma_i$ are the Pauli spin matrices, generate the
{\it Pauli group} using matrix multiplication.  The elements of this
group are known as Pauli operators and provide a basis of unitary
operators on ${\cal H}_2$.

Just as a qubit is realized as a state in a Hilbert space of dimension
two, a qudit is realized as a state in a $d$--dimensional Hilbert
space ${\cal H}_d$.  It is useful to choose a computational basis
$\{ |s\rangle; \, s=0,1,\ldots,d-1 \}$ for ${\cal H}_d$, which
serves as the generalization of the binary basis $\{ |0\rangle, \,
|1\rangle \}$ of the qubit.

A basis for unitary operators on ${\cal H}_d$ is given by the
{\it generalized Pauli operators}~\cite{Pat88,Kni96,Got00}
\begin{equation}
  \label{eq:PauliOperators}
  (X_d)^a (Z_d)^b , \quad a,b \in 0,1,\ldots d-1 \, ,
\end{equation}
where $X_d$ and $Z_d$ are defined by their action on the computational
basis as follows:
\begin{eqnarray}
  \label{eq:ActionOfXOnCompBasis}
  X_d | s \rangle &=& | s+1\ ({\rm mod}\ d) \rangle \, , \\
  \label{eq:ActionOfZOnCompBasis}
  Z_d | s \rangle &=& \exp (2 \pi {\rm i} s/d) | s \rangle \, . 
\end{eqnarray}
The operators $X_d$ and $Z_d$ generate a group under matrix
multiplication, known as the {\it generalized Pauli group}.  Note
that $X_d$ and $Z_d$ are non--commutative and obey
\begin{equation}
  \label{eq:NonCommutivityOfXZ}
  Z_d X_d = \exp(2\pi {\it i}/d) X_d Z_d \, .
\end{equation}
In the following, we give a representation of $X_d$ and $Z_d$ in the
$d$--dimensional Hilbert space of a SU(2) irrep of highest weight
(angular momentum) $j= (d-1)/2$.  The relevant generalized Pauli
operators can be viewed in terms of SU(2) angular momentum and phase
operators~\cite{Vou90}.

Consider the standard basis for the su(2) algebra $\{ \hat{J}_z,
\hat{J}_\pm = \hat{J}_x \pm {\rm i} \hat{J}_y \}$.  Let $\{ |j,m)_z;
\, m=-j,\ldots, j \}$ denote the standard weight basis for the
Hilbert space ${\cal H}_{d=2j+1}$ for an SU(2) irrep of highest
weight (angular momentum) $j$.  We use the simplifying notation
of Vourdas~\cite{Vou90} where we allow $m$ to take all the integer (or
half--integer) values modulo $2j+1$, thus defining $|j,j+1)_z =
|j,-j)_z$.

With the computational basis defined to be
\begin{equation}
  \label{eq:CompBasisSU(2)WeightBasis}
  |s\rangle \equiv |j, j-s)_z \, , \quad s=0,1,\ldots, d-1\, ,
\end{equation}
we now write the generators of the generalized Pauli group in terms of
operators that act in a natural way on our SU(2) basis states.
Because the basis states are eigenstates of $\hat{J}_z$, we have
\begin{equation}
  \label{eq:su(2)RepOfZ}
  Z_d \mapsto \exp \bigl( 2\pi {\rm i} (j - \hat{J}_z)/d \bigr) \, ,
\end{equation}
which is unitary and satisfies Eq.~(\ref{eq:ActionOfZOnCompBasis}).
For the generalized Pauli operator $X_d$, we use 
\begin{equation}
  \label{eq:su(2)RepOfX}
  X_d \mapsto \sum_{m=-j}^j |j,m)_z (j,m+1|\, .
\end{equation}
One can easily check that $X_d$ satisfies
Eq.~(\ref{eq:ActionOfXOnCompBasis}) under the identification of
Eq.~(\ref{eq:su(2)RepOfX}) and that it is unitary.  The
operators $X_d$ and $Z_d$ satisfy Eq.~(\ref{eq:NonCommutivityOfXZ}) and
together generate a representation of the generalized Pauli group for
a qudit. It is convenient to view $X_d$ as
the exponent of a Hermitian operator $X_d = \exp (2\pi {\rm i}\,
\hat{\theta}_z/d)$, just as $Z$ is generated by the operator
$\hat{J}_z$.  The operator $\hat{\theta}_z$ is the SU(2) phase operator
of Vourdas~\cite{Vou90}.

It is also possible to realize the operators $X_d$ and $Z_d$ as
operators that act naturally on the space ${\Bbb H}_d$ of dimension
$d$ spanned by harmonic oscillator states of no more than $d-1$
bosons.  We define the computational basis to be the set of harmonic
oscillator energy eigenstates
\begin{equation}
  \label{eq:CompBasisOscillatorEigenstates}
  |s\rangle \equiv |n = s \rangle_{{\rm HO}} \, , \quad s=0,1,\ldots,d-1
   \, ,
\end{equation}
where $\hat{N}|n\rangle_{{\rm HO}} = n|n\rangle_{{\rm HO}}$.  Again,
we apply the cyclic notation $|d\rangle = |0\rangle$.  Encoding a
qudit in an oscillator is important not only for investigating the
$d\to \infty$ limit, but also for realizing a qudit experimentally and
for creating error--correcting codes for qudit--based
computation~\cite{Got00}.

In this Hilbert space, the generators $X_d$ and $Z_d$ of the generalized
Pauli group for a qudit become
\begin{equation}
  \label{eq:HPrepOfXandZ}
  X_d \mapsto \sum_{s=0}^{d-1} |s+1\rangle \langle s| \, , \quad
  Z_d \mapsto \exp( 2 \pi {\rm i} \hat{N} /d) \, ,
\end{equation}
which are unitary on ${\Bbb H}_d$.  It is convenient to view $X_d$
as the exponent of a Hermitian operator $\hat{\theta}_z$, such that
$X_d = \exp (2\pi {\rm i}\, \hat{\theta}_z/d)$; the operator
$\hat{\theta}_z$ is the Pegg--Barnett phase operator~\cite{Peg97}.  We
will call this representation of the generalized Pauli group the
{\it number representation}.

The explicit realization of $X_d$ and $Z_d$ as unitary operators on
the harmonic oscillator Hilbert space enables us to investigate the $d
\to \infty$ limit; the limiting procedure for phase operators has been
thoroughly investigated~\cite{Peg97,Lyn95}.  In this limit, the
computational basis remains the harmonic oscillator energy eigenstates
(now including all states $s = 0,1,\ldots,\infty$), following
Eq.~(\ref{eq:CompBasisOscillatorEigenstates}).  It is natural to
generalize the operator $X_d$ to a continuous transformation $X(x)$,
generated by the phase operator $\hat{\theta}_z$; i.e.,
\begin{equation}
  \label{eq:CVX(x)}
  X(x) \equiv \exp({\rm i}x \, \hat{\theta}_z) \, ,\quad x \in
  {\Bbb R}\, . 
\end{equation}
Similarly, the CV generalization of $Z_d$ is obtained by replacing the
finite angle $2\pi /d$ in the expression for $Z_d$ in
Eq.~(\ref{eq:HPrepOfXandZ}) by the continuous angle $z \in
{\Bbb R}$, so that we now have the unitary transformation $Z(z)$
defined by
\begin{equation}
  \label{eq:CVZ(z)}
  Z(z) \equiv \exp({\rm i}z \hat{N}) \, .
\end{equation}

Note that, in extending the above representation of the generalized
Pauli group from qudits to CV representations, we do not obtain the
usual generalization as the Heisenberg--Weyl group, with position
$\hat{x}$ and momentum $\hat{p}$ operators as generators.  Instead,
the generalized Pauli operators are generated by the number operator
$\hat{N}$ and the phase operator $\hat{\theta}_z$; these operators are
in a sense ``conjugate'' like momentum and position, but there exist
challenging problems with defining the phase operator on the
infinite--dimensional Hilbert space ${\Bbb H}_\infty$ of the
harmonic oscillator~\cite{Peg97,Lyn95}.  It is also interesting to
note that the states of the computational basis for the limiting case
remain harmonic oscillator energy eigenstates, not position (or
momentum) eigenstates or squeezed Gaussians as are commonly used for
CV quantum computing.  In what follows, we will refer to this
representation as the {\it number representation of the generalized
  CV Pauli group}.

It is possible to construct another realization of $X_d$ and $Z_d$ in
the Hilbert space ${\cal H}_d$ for an irrep of SU(2) where the
computational basis is given by SU(2) phase states.  This
representation is ``dual'' to the number representation.  Consider the
relation ${\rm i} X_2 = \exp ( {\rm i}(\pi/2)X_2 )$ for qubits;
i.e., that
\begin{equation}
  \label{eq:OneIsRotationOfZero}
  |1\rangle = X_2 |0\rangle = (-{\rm i})e^{{\rm i}(\pi/2)X_2}
   |0\rangle \, .
\end{equation}
The Pauli operator $X_2$ has two interpretations, each of
which can be generalized in a different way.  In the number
representation, we interpreted $X_2$ as a number state raising operator
$|1\rangle = X_2 |0\rangle$ and generalized this operator as such.
However, using the relation~(\ref{eq:OneIsRotationOfZero}), we can
also view $X_2$ as a rotation.  (Using the su(2) representation $X_2 =
2\hat{J}_x$, this rotation is about the $x$--axis.)  Thus, the state
$|1\rangle$ is obtained (up to a phase) by rotating $|0\rangle$ by an
angle $\pi$ about the $x$--axis.  The computational basis states
needed for this type of generalization to qudits are ``SU(2) phase
states'' and have been investigated by Vourdas~\cite{Vou90} (although
using rotations generated by $\hat{J}_z$ rather than $\hat{J}_x$).
These states form an orthonormal basis for the SU(2) irrep and are
``dual'' to the usual weight basis.

Let $\{ |j,m)_x ;\, m=-j,\ldots,j \}$ be the weight basis for an SU(2)
irrep of angular momentum $j = (d-1)/2$, where $\hat{J}_x$
rather than $\hat{J}_z$ is diagonal; i.e., $\hat{J}_x|j,m)_x
= m|j,m)_x$.  For this representation, we define the computational
basis states to be
\begin{eqnarray}
  \label{eq:PhaseStateCompBasis}
  |s\rangle &\equiv&
  \frac{1}{\sqrt{d}} \sum_{m=-j}^j \exp ( 2 \pi {\rm i} ms/d )
  |j,m)_x\, , \quad d\ {\rm odd,} \\
  |s\rangle &\equiv&
  \frac{1}{\sqrt{d}} \sum_{m=-j}^j \exp ( 2 \pi {\rm i}
  (m+\frac{1}{2}) s/d ) |j,m)_x \, , \quad d\ {\rm even.}
\end{eqnarray}
These states form an orthonormal basis for ${\cal H}_d$~\cite{Vou90}.
They are referred to as SU(2) {\it phase states} because they are
eigenstates of the SU(2) phase operator, defined below.

The generalized Pauli operator $X_d$ on this computational basis is
given by
\begin{eqnarray}
  \label{eq:RotationQuditX}
  X_d &\mapsto& 
  \exp \bigl( 2 \pi {\rm i} \hat{J}_x / d \bigr)\, , \quad d\ {\rm odd,} \\
  X_d &\mapsto& \exp ( -{\rm i}\pi/d ) \exp \bigl( 2 \pi {\rm i}
  \hat{J}_x / d \bigr)\, , \quad d\ {\rm even,} 
\end{eqnarray}
satisfying Eq.~(\ref{eq:ActionOfXOnCompBasis}).  Note that $(X_d)^d =
\hat{\openone}$ for both $j$ integral and half--integral (i.e.,
spinor).  The generalized Pauli operator $Z_d$ is given by
\begin{equation}
  \label{eq:RotationQuditZ}
  Z_d \mapsto \sum_{s=0}^{d-1} \exp(2\pi{\rm i}s/d) |s\rangle\langle
  s| \, ,
\end{equation}
which is unitary and satisfies Eq.~(\ref{eq:ActionOfZOnCompBasis}).
Note that we can express $Z_d$ as the exponent of a Hermitian operator,
\begin{equation}
  \label{eq:RotationQuditXAsExp}
  Z_d = \exp \bigl(2 \pi {\rm i}\, \hat{\theta}_x /d \bigr) \, , \quad 
  \hat{\theta}_x \equiv \sum_{s=0}^{d-1} s |s\rangle\langle s| \, ;
\end{equation}
the operator $\hat{\theta}_x$ is the SU(2) phase operator.

Note that this representation of the generalized Pauli group is
``dual'' to the number representation in the same sense that the
position and momentum representations of the harmonic oscillator are
dual.  For the number representation, the computational basis states
are eigenstates of $\hat{J}_z$, and the phase operator
$\hat{\theta}_z$ generates the ``ladder'' transformations.  In the
{\it phase representation} given here, the computational basis states
are eigenstates of the phase operator $\hat{\theta}_x$, i.e., ``phase
eigenstates'', and it is $\hat{J}_x$ which generates the ladder
transformations via rotations about the $x$--axis.  Both of these
representations can be considered natural generalizations of the qubit
case, because the standard computational basis $|0\rangle =
|\frac{1}{2},\frac{1}{2})_z$ and $|1\rangle =
|\frac{1}{2},-\frac{1}{2})_z$ are both eigenstates of $\hat{J}_z$
and phase eigenstates of $\hat{\theta}_x$.

As with the number representation, this phase representation of the
generalized Pauli group can be expressed in a harmonic oscillator
Hilbert space.  Again, the $d\to\infty$ limit yields challenging
problems: it is well known that phase eigenstates do not exist in the
infinite--dimensional Hilbert space ${\Bbb H}_\infty$ of the
harmonic oscillator~\cite{Peg97}.

In any experimental realization, the problems associated with taking
the $d\to \infty$ limit would not arise.  A physically realistic
system would have a finite energy cutoff (and an associated resolution
in time and thus phase), and so experimental CV computation would in
actuality involve qudits with finite (although possibly very large)
$d$.  As a result of our well--defined limiting procedure for qudits,
the above realization of CV quantum computation is applicable to such
a physically realistic system.

TO perform universal CV computation~\cite{Llo99}, it is necessary to
be able to realize an arbitrary unitary transformation on a single
qudit, and to have a controlled two--qudit interaction gate such as
the SUM gate~\cite{Got00}.  Considering an optical realization, an
arbitrary unitary transformation on a single qudit, to any desired
precision, can be performed {\it efficiently} using a combination of
linear optics, parametric down--conversion, and a nonlinear optical
Kerr medium~\cite{Llo99}.  By this combination, one can approximate
(to arbitrary accuracy) any polynomial Hamiltonian in
$\hat{a}^\dagger$ and $\hat{a}$.  Of particular importance is to
realize the Fourier transform operation on a single qudit, which takes
number eigenstates to phase eigenstates and vice versa.  This
operation is the generalization of the Hadamard transformation for
qubits.

For quantum computation, we must also realize a gate that performs a
two--qudit interaction.  Consider two oscillators coupled by the
four--wave mixing interaction Hamiltonian $\chi \hat{N}_1 \hat{N}_2 =
\chi \hat{a}_1^\dagger \hat{a}_1 \hat{a}_2^\dagger \hat{a}_2$.  This
Hamiltonian for an optical system describes a four--wave mixing
process in which $\chi$ is proportional to the third--order nonlinear
susceptibility~\cite{Mil83}.  Let oscillator $1$ be in a state $|s_1
\rangle_1$ encoded in the number state basis, and let oscillator $2$
be in a state $|s_2 \rangle_2$ encoded in the phase state basis.  This
interaction Hamiltonian generates the transformation
\begin{equation}
  \label{eq:SUMHamiltonian}
  e^{-{\rm i} \chi \hat{N}_1 \hat{N}_2 t} |s_1 \rangle_1 \otimes |s_2
  \rangle_2 = |s_1 \rangle_1 \otimes |(\frac{\chi t}{2\pi}) s_1 + s_2
  \rangle_2 \, .
\end{equation}
Thus, with time $t=2\pi\chi^{-1}$, this Hamiltonian generates the SUM
transformation $|s_1\rangle_1\otimes|s_2\rangle_2 \to
|s_1\rangle_1\otimes|s_1+s_2\rangle_2$.

In summary, we have presented a new form of continuous variable
computation in terms of number and phase operators.  This new approach
has the advantage over position--eigenstate CV computation in that the
computational basis states, for large but finite $d$, are
well--defined and obtainable, and do not require
``infinite--squeezing'' of Gaussian wavepackets.

This project has been supported by an Australian Research Council
Large Grant and by a Macquarie University Research Grant.  We
acknowledge helpful discussions with Samuel Braunstein.

\end{document}